# Why a diffusing single-molecule can be detected in few minutes by a large capturing bioelectronic interface

*Eleonora Macchia‡, Liberato De Caro‡, Fabrizio Torricelli, Cinzia Di Franco, Giuseppe Felice Mangiatordi, Gaetano Scamarcio\* and Luisa Torsi\**

**Dr. Eleonora Macchia**
Faculty of Science and Engineering, Åbo Akademi University, 20500 Turku (Fi)

Dr. Liberato De Caro, Dr. Giuseppe Felice Mangiatordi
Institute of Crystallography, National Research Council, via Amendola 122/O, 70126 Bari (I)

Prof. Fabrizio Torricelli
Dipartimento Ingegneria dell'Informazione, Università degli Studi di Brescia, via Branze 38, 25123 Brescia (I)

**Dr. Cinzia Di Franco, Prof. Luisa Torsi**
Dipartimento di Chimica, Università degli Studi di Bari "Aldo Moro", 70125 Bari (I)

**Dr. Cinzia Di Franco, Prof. Gaetano Scamarcio**
CNR, Istituto di Fotonica e Nanotecnologie, Sede di Bari, 70125 Bari (I)

Prof. Gaetano Scamarcio
Dipartimento Interateneo di Fisica "M. Merlin", Università degli Studi di Bari "Aldo Moro", 70125 Bari (I)

Dr. Elenora Macchia, Prof. Fabrizio Torricelli, Dr. Cinzia Di Franco, Prof. Gaetano Scamarcio, Prof. Luisa Torsi
CSGI (Centre for Colloid and Surface Science), 70125 Bari (I)

‡ Equally contributing authors
\* gaetano.scamarcio@uniba.it, luisa.torsi@uniba.it

**Abstract:** Single-molecule detection at a nanometric interface in a femtomolar solution, can take weeks as the encounter rate between the diffusing molecule to be detected and the transducing nano-device is negligibly small. On the other hand, several experiments prove that macroscopic label-free sensors based on field-effect-transistors (FETs), engaging micrometric or millimetric detecting interfaces are capable to assay a single-molecule in a large volume within few minutes. The present work demonstrates why at least a single molecule out of a few diffusing in a 100 µl volume has a very high probability to hit a large capturing and detecting electronic interface. To this end, sensing data, measured with an electrolyte-gated (EG-)FET whose gate is functionalized with $10^{12}$ capturing anti-immunoglobulin G, are here provided along with a Brownian diffusion-based modelling. The EG-FET assays solutions down to some tens of zM in concentrations with volumes ranging from 25 µl to 1 ml in which the functionalized gates are incubated for times ranging from 30 s to 20 min. The high level of accordance between the experimental data and a model based on the Einstein's diffusion-theory proves how the single-molecule detection process at large-capturing interfaces is controlled by Brownian diffusion and yet is highly probable and fast.

## 1. Introduction

The assay of biomarkers with a sensing system endowed with single-molecule limit-of-detection (LOD),[1] can be a game changer in enabling early detection of progressive diseases. For many years nanotechnologies seemed to offer feasible solutions by means of the so called *near-field* approach at nanoscopic interfaces.[2] However, the difference between single-molecule resolution and single-molecule LOD should be kept in mind. When a target molecule is dispersed in a volume of 10 - 100 µl comprising also the nanometric detecting interface, the interaction cross-section between the two is extremely unlikely. This is known as the *diffusion-barrier* issue,[3],[4] which can be exemplified by saying that the probability for the interaction of a nanometric interface and a target molecule driven by diffusion is not negligible only within a volume of ~ 1 µm³ (1 femtoliter, fl). Note that 1 particle in 1 fl corresponds to a nanomolar (nM) solution. As an instance, a simulation study shows that a timescale of several days are needed for ten molecules, out of $10^6$ in 100 µl, *i.e.* a femtomolar (fM) concentration, to hit a at a nanometric interface immersed in the same volume.[4] Experimental evidences have been gathered on many nano-transducers,[5] from nanopores[6] to nano-transistors,[7] proving how they can detect with single-molecule resolution, but only at LODs larger than picomolar (pM) concentration. Hence, nano-interfaces while being extremely effective to study rarer events offering indeed single-molecule resolution but not low LOD, cannot be used to assay at extremely low concentration.

*Large-area* or *wide-field*[2] transducing interfaces, exposing a much larger active area can be a viable solution. However, it is often assumed that alike nano-transducing interfaces, also large-area ones cannot detect a single molecule in a large-volume (*e.g.*, 10 - 100 µl), because affected by the *diffusion-barrier* issue too. This is proven erroneous by several published experimental pieces of evidence, involving mostly field-effect transistor (FET) detections[8] at a large-area interface. As an instance, a bioelectronic sensor based on a AlGaN/GaN high electron-mobility transistor was proposed to detect antigen proteins such as Human Immunodeficiency Virus-1 Reverse Transcriptase, Carcinoembryonic Antigen, N-terminal pro b-type natriuretic peptide, and C-reactive protein (CRP), even in human serum at the fM concentration level.[9] The gate surface, biofunctionalized with capturing antibodies, is 100-µm-wide and the assay was completed in 5 minutes with the lowest detections in the fM range. Other highly performing bioelectronic sensors are the FETs gated *via* an ionically-conducting and electronically-insulating electrolyte, known as Electrolyte-gated- (EG)-FETs,[10] foreseen to be produced by scalable large-area, low-cost approaches.[11] These sensors[12],[13] are endowed with selectivity *via* the bio-functionalization of a millimeter-wide sensing interface with a high density (up to $10^{12}$/cm$^2$) of recognition elements.[14],[15],[16] An EG-FET sensor with a graphene channel bearing $10^{11}$ cm$^{-2}$ human olfactory receptors, can selectively binding the myl-butyrate odorant marker down to a LOD of 40 attomolar (aM, $10^{-18}$ M) with a response-time shorter than 1 second.[17] Likewise, a graphene based EG-FET was shown able to detect Anthrax Toxin at a LOD of 12 aM, in 200 seconds.[18] More recently the LOD was reduced down to tens of zeptomolar (zM) with the *Single-Molecule assay with a large-Transistor* (SiMoT) technology, involving an organic semiconductor based EG-FETs.[16] This is a single-molecule assay as a 100 µl of a 10 - 20 zM solution comprises 1 ± 1 molecules. Also in this case $10^{12}$ recognition elements were covalently attached at a millimeter-wide gate (area of 0.2 cm$^2$) electrode and detections were possible after 10 minutes of incubation in the solution to be assayed.[19] Hence, SiMoT sets in 2018 a world record in label-free single-molecule detection and relevantly it was demonstrated to uniquely detect both proteins (CRP, IgG, IgM),[20],[21],[16] including virus capsids' one (HIV-p24),[22],[23] aptamer[24] and genomic marker[25] also in serum. Lately, single-molecule chiral[26] as well as CoV-SERS-2 virus[27] detection was proposed with large-area, fast-responding *ad hoc* functionalized organic FETs. The elicited pieces of evidence gathered on completely different FET structures by several research groups, demonstrate that a single molecule in 100 µl (concentration of ~ 10 - 20 zM) can diffuse and eventually impinge in the minute timescale, on a millimeter-wide (*e.g.*, ~ 0.2 cm$^2$) surface populated with trillions of recognition elements. The binding generates a signal that equals the noise average-level plus three times its standard deviation (LOD definition).[1]

The present work undertakes a systematic investigation to explain why when few molecules (< 10) diffuse in a large volume (*e.g.*, 100 µl) comprising also a millimeter-wide detecting interface, within 10 minutes at least one of them impinges on the large-interface generating a detectable signal at the LOD. The engaged interface is the millimeter-wide gate of a SiMoT EG-FET device, covered by trillions of immunoglobulin G (anti-IgG) capturing antibodies, while the target molecule is the IgG affinity antigen. The response data, acquired assaying solutions down to 60 ± 30 zM encompassing different volumes (25 µl - 1 ml) and incubation times (30 sec – 20 min), were successfully modeled with the Einstein's diffusion-theory.

## 2. Results and discussion

In **Figure 1a** a typical EG-FET SiMoT device structure is sketched. It includes a poly(3-hexylthiophene-2,5-diyl) - P3HT FET channel with the source (S) and drain (D) interdigitated electrodes, along with a 5 mm-diameter Au gate functionalized with a grafted layer of $10^{12}$ anti-IgG capturing antibodies[16] addressed as *sensing gate*. A bare-gold electrode, having the same size and serving as *reference gate* is also present. A well, glued around the channel, is filled with deionized water. This is addressed as the *measuring well* and the sensing and the reference gate immersed into it are, alternatively, capacitively coupled to the P3HT channel. The latter and the reference gate are always in the measuring well. The reference gate enables to control the level of the current flowing in the channel, at every stage of the sensing assay. Conversely, the sensing gate is alternatively immersed into the *measuring* well and into a separate one, addressed as *incubation* well, that contains the solutions to be assayed. The solutions to be assayed are based on phosphate buffered saline (PBS) so as to mimic a real fluid physiological high ionic strength of 162 mM and pH of 7.4. They are spiked with a given number of IgGs. The necessity to separate the *measuring* from the *incubation* wells, is dictated by the need of measuring, while not screened, the electrostatic changes elicited by the IgG /anti-IgG binding. Upon binding the sensing gate capacitively coupled to the FET channel undergoes a work-function shift, measured as a change of the FET current. In deionized water, the Debye length is maximized and so is the FET current change.[16]

The *incubation* well is filled with IgG standard solutions, whose volume is 100 µl or 1 ml. A schematic diagram of the incubation process in these larger volumes is given in **Figure 1b**.

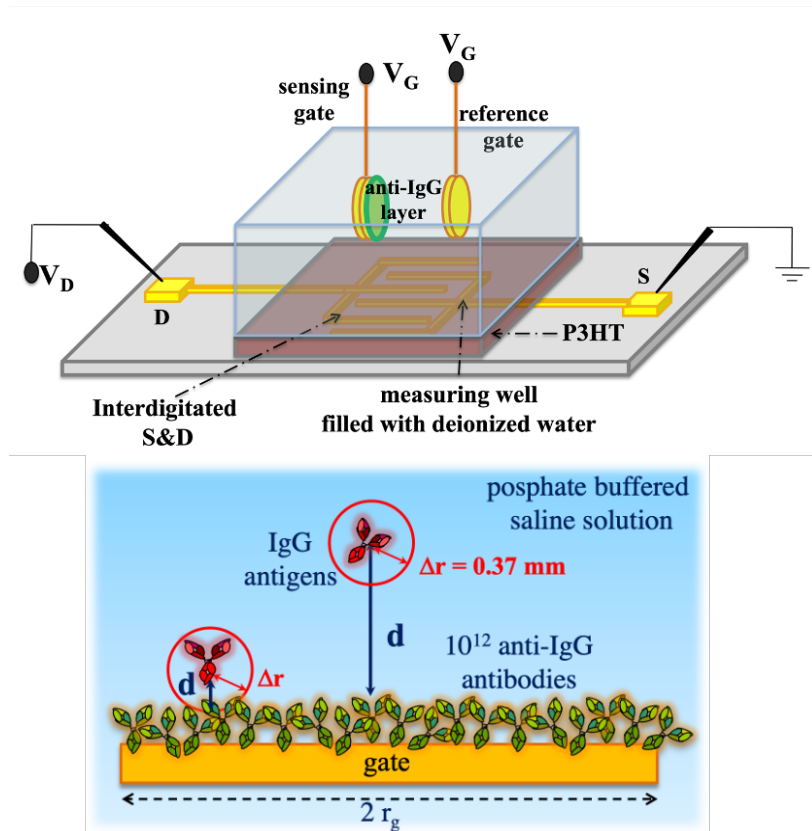

**Figure 1**. *(a) EG-FET SiMoT set-up comprising a P3HT channel coupled to either one of the two gates immersed in the well. (b) Schematic cross-sectional view of the incubation step carried out in the solution to be assayed (volumes of 100 µl or 1 ml) in contact with the functionalized sensing gate surface of radius $r_g$. An IgG antigen can randomly move, in a time $\Delta t$, within a sphere of radius $\Delta r$ being 0.37 mm in $\Delta t = 600$ s.*

The *sensing* gate, whose radius ($r_g$) is smaller than the edge of the *incubation* well (not shown), is depicted along with few IgG target antigens.

In **Figure 2** the normalized electronic responses of the SiMoT EG-FET devices are plot as a function of the total number of molecules, $N$, in the assayed volumes. The response is evaluated as the negative relative current shift upon sensing, namely $(-\Delta I/I_0) = [-(I - I_0)/I_0]$, where $I_0$ is the base-line source-drain current measured after incubating the anti-IgG biofunctionalized sensing gate in a PBS solution where no IgG molecules are present. The current I is measured after incubating the same gate in the solutions to be assayed encompassing a given number, N, of IgG molecules in a given volume. The saturated value of $(-\Delta I/I_0)$, $(-\Delta I/I_0)_{sat}$ is sample dependent as it is related to the quantity of available anti-IgG binding sites.[20] As anticipated, the assayed samples are high ionic strength PBS standard solutions of the IgG antigens. The 100 µl of the PBS solution comprising $4 \pm 2$ IgG molecules where both the Poisson sampling and the concentration errors are evaluated at first.[16] The same gate is incubated, afterwards, in 100 µl solutions encompassing $N = 39 \pm 6$, $N = 392 \pm 20$ and $N = 3.92 \cdot 10^3 \pm 60$, $N = 3.92 \cdot 10^4 \pm 2 \cdot 10^2$, $N = 3.92 \cdot 10^5 \pm 6 \cdot 10^2$, $N = 3.92 \cdot 10^6 \pm 2 \cdot 10^3$ and $N = 3.92 \cdot 10^7 \pm 6 \cdot 10^3$ IgG molecules, respectively. The data, given as red circles, are the average over at least two replicates (using two different anti-IgG biofunctionalized gates) while the reproducibility error bars are given as one standard deviation. A similar dose curve is measured incubating anti-IgG gates in a 1 ml (blue triangles) of PBS standard solutions. The minimum concentration assayed is $60 \pm 30$ zM and the maximum is $600 \pm 2$ fM. Relevantly, the encounter between the antigen and the capturing gate surface occurs during the incubation in the PBS standard solutions when no bias is applied so, no field induced drifting contributes to the antigen motion. The data for the incubation carried out in a smaller volume of 25 µl are given in **Figure S1** (Supplementary Information, **SI - Section 1**). In this case a droplet was deposited on the gate as it was not possible to immerse the whole gate into such a small volume.

As customary for the SiMoT sensing protocol, the level of the source-drain current in the channel induced by a bare gold gate always kept in the measuring cell, is checked before and after the measurement of each dose curve, to control that its relative variation is within 5 %.[16],[19],[20],[21],[22] A negative control experiment was also performed by exposing an anti-IgG functionalized gate to PBS standard solutions of the Immunoglobulin M (IgM) in the same concentration range span in the IgG assays. As IgM does not bind to anti-IgG (**Figure S2, SI - Section 2**), these data are taken as the noise, whose average level plus three-times its standard deviation results in a LOD level of 22%, marking the lowest acceptable response at a confidence level of 99%.[1]

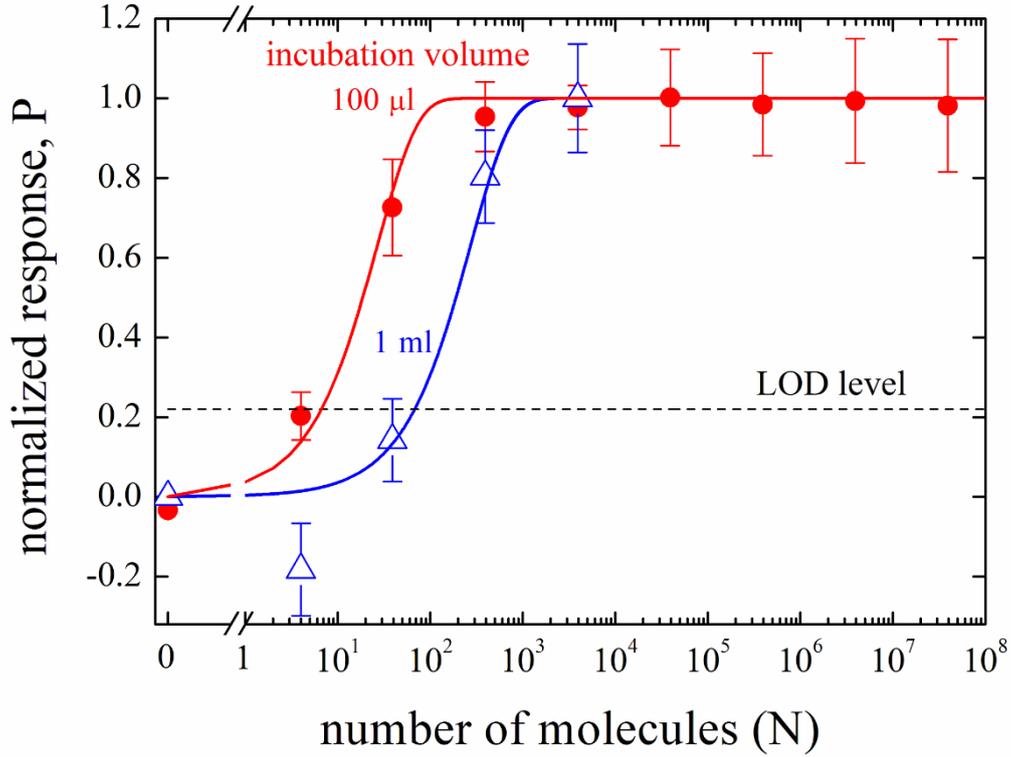

***Figure 2.*** *Normalized SiMoT EG-FET responses, $(\Delta I/I_0)/(\Delta I/I_0)^{sat}$ with $(-\Delta I/I_0)^{sat} = 0.74 \pm 0.13$ for the data taken in 100 µl and $(-\Delta I/I_0)^{sat} = 0.47 \pm 0.14$ for the data measured in 1 ml. The anti-IgG functionalized gates were incubated for 10 minutes (600 s) into 100 µl (red symbols) and 1 ml (blue symbols) solutions of N IgGs, with N ranging from $4 \pm 2$ to $3.92\ 10^7 \pm 6 \cdot 10^3$. Error bars are relevant to the reproducibility error indicated as one standard deviation over at least two replicates. On the y-axis the P probability function (vide infra) is also given, and the solid lines are the result of the modelling. The black dotted line sets the level of the LOD.*

The solid curves in **Figure 2** are calculated assuming a Brownian motion for the diffusion of the IgG molecules, undergoing stochastic collisions with the solvent molecules that are much smaller in mass and size.[28] The Einstein's theory of diffusion relates the IgGs diffusion coefficient D to their mean squared displacement $\langle\Delta r^2\rangle = \langle\Delta x^2\rangle + \langle\Delta y^2\rangle + \langle\Delta z^2\rangle$, with:

$$\Delta r = (\langle\Delta r^2\rangle)^{1/2} = (6 D \cdot \Delta t)^{1/2} \qquad (1)$$

$D = k_B \cdot T / (6 \pi \cdot \eta \cdot a)$, $k_B$ being the Boltzmann constant, T the absolute temperature, $\eta$ the solvent viscosity and *a* the hydrodynamic radius of the diffusing Brownian protein. In **SI – Section 3** further details on the model are provided. The diffusion coefficient used, $D = 3.89 \cdot$ $10^{-7}$ cm²/s, is extracted from the analysis of photon-correlation spectroscopy experiments for IgG monomers.[29] Using the above expressions it can be estimated that an IgG undergoing a Brownian motion for Δt = 600 s (10 min) will span a spherical volume with a radius Δr ~ 0.037 cm. The portion of the solution that is close enough to the gate surface ($d \leq \Delta r$, see **Figure 1b**) to enable an antigen-antibody interaction within 600 s can be approximated by a volume of $V_{\Delta r}$, based on a cylindrical disk of radius $r_g$ and height 0.5 · Δr. The factor 0.5 accounts for the packing of a sphere of radius Δr into a cube of edge 2 · Δr. Hence, $V_{\Delta r}$ is given by:

$$V_{\Delta r} = \pi \cdot \Delta r \cdot 0.5 \cdot r^2_g \qquad (2).$$

One further aspect of the Brownian motion model worth commenting concerns the IgG rotational displacement angle for which equations and constrains equivalent to those given for the translational displacement hold. The time an IgG takes to span the whole solid angle 4π can be computed to be just 25 μs (**SI - Section 3**). Brownian motion, thus, also enables the antigen to quickly find the right orientation to correctly bind a given antibody, even when the latter is not perfectly oriented towards the solution to be assayed.

The fraction *f* of the incubating volume V that is close enough to the gate surface to enable the antigen-antibody interaction within 600 s, is:

$$f = (V_{\Delta r} / V) = (\pi \cdot 0.5 \cdot \Delta r \cdot r^2_g) \cdot / V \qquad (3)$$

Plugging **Equation 1** into **Equation 3** the following results:

$$f = \left[ \frac{\pi}{2} \sqrt{6D\,\Delta t} \cdot r_g^2 / V \right] \qquad (4).$$

The SiMoT EG-FET response can be elicited by just one antigen being captured. [16],[11],[20] To model this occurrence, assuming that *N* IgG antigens are randomly dispersed in the assayed volume V, *f* is defined as the probability that one of them happens to be sufficiently close to the gate surface (namely, within a Δr distance), to eventually collide on it. The (*1 - f*) term is the probability that no antigen is in the *f* portion of the volume closer to the gate surface. Hence, the conditional probability that at least one antigen, out of N, is in the proximity by the gate surface, is given by:

$$P = f + (1-f) \cdot f + (1-f)^2 \cdot f + \ldots (1-f)^{N-1} \cdot f \quad (5).$$

The rationale behind **Equation 5** is to model a system of N diffusing IgGs by the conditional probability function P built as follows: one IgG out of the N present in the assayed volume V, holds a probability $f$ to find itself in the $V_{\Delta r}$ fraction of the whole volume (**Equation 3**). This IgG will be, hence, close enough to the gate to hit against the anti-IgG functionalized surface (**Figure 1b**) within a $\Delta t$ of 10 minutes. At this point the binding can easily take place as the IgG spans, rotating, the whole solid angle in only 25 µs. Thus, the antigen can quickly find the right orientation to bind the anti-IgG bumped into, out of the $10^{12}$ antibodies populating the gate surface. According to the hypothesis, a second IgG can be wherever, but in the volume $V_{\Delta r}$. So, the second term in **Equation 5**, $(1-f) \cdot f$, expresses the probability that no second IgG is in the $f$ portion of the volume closer to the gate surface. A third IgG will also not be in the $f$ fraction and the third term will be $(1-f)^2 \cdot f$; and so on for the other N-4 particles populating the volume. Relevantly **Equation 5** encompasses only independently measured physical parameters such as the $\Delta t$ incubation time, the volume V assayed, the diffusing Brownian coefficient D for an IgG monomer.[29]. The probability P is given as a function of the incubation volume encompassing different N values in **Figure S3 (SI - Section 4)**. The complete expression of **Equation 5** is provided in **SI – Section 4** along with the list of the physical quantities and constants used for the calculations (**Table 1S**).

The plot of the *P* probability function (**Equation 5**) results in the solid curves given in **Figure 2**. Indeed, within one standard deviation all the experimental trends, are reproduced by the model. The very good agreement between the experimental data reported in **Figure 2** and **Equation 5** is further proven by the chi-squared test (**SI – Section 5**) assessing a very high chance of 95% that such experimental data are successfully predicted by the P' modeling function. The modelling curves serves also to assess the number of molecules at the LOD level, $N_{LOD}$, being $8 \pm 3$ for the 100 μl data and $175 \pm 13$ the 1 ml ones.

To deepen the understanding of the phenomenon investigated, the SiMBiT EG-FET $\Delta I/I_0$ responses measured at different incubation times, namely $\Delta t$ of 30 s, 1 min, 5 min, 10 min and 20 min, are modelled. To this end, the so far used anti-IgG functionalized gates are engaged to assay a 100 μl solution of $N = 39 \pm 6$ IgG molecules.

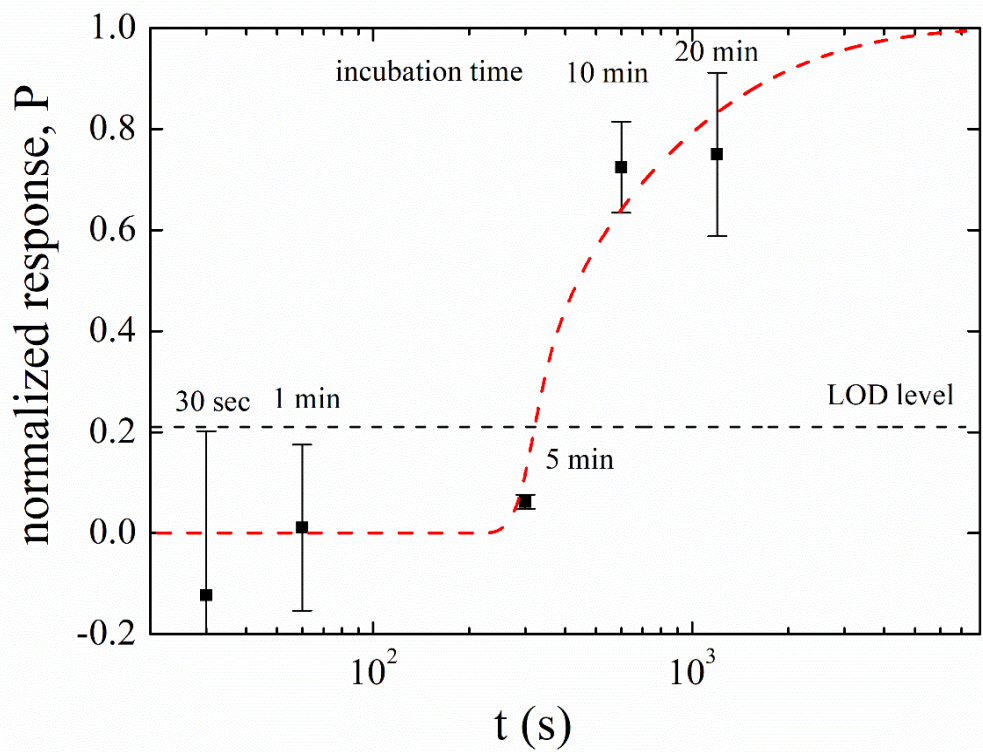

**Figure 3.** *Normalized SiMoT EG-FET responses $(\Delta I/I_0)/(\Delta I/I_0)^{sat}$ with $(- \Delta I/I_0)^{sat} = 0.74 \pm 0.13$. Anti-IgG functionalized gates incubated for $\Delta t$ 30 seconds, 1, 5, 10 and 20 minutes into a 100 μl volume containing a $39 \pm 6$ zM IgGs. Error bars are the reproducibility standard deviations over two replicates. On the y-axis the P probability function is also given, and a solid red line is the result of the modelling; the dotted portion of the red line is relevant to the timeframe ($\Delta t < 250$ s) in which the P functions does not return a physically meaningful value. The black dotted line sets the level of the LOD.*

The data are plotted in **Figure 3** as the normalized $\Delta I/I_0$ *vs.* the incubation time $\Delta t$ along with the y-axis error bars taken as one standard deviation over two replicates. In this case 10 different solutions are assayed with an equal number of different gates. As it is apparent an incubation time Δt between 5 - 10 minutes is needed to see a response beyond the LOD level. The modelling is carried out considering the very same function P given by **Equation 5**, that can indeed model the (- $\Delta I/I_0$) response as a function of the incubation time Δ*t*, because it implicitly depends on the square-root of the time *via* the Stokes-Einstein relation (**Equation 1)**. The P probability function (red line in **Figure 3**) reproduces very well the experimental data as assessed, also in this case, by the chi-squared test (**SI – Section 5**). **Equation 5** returns no response for Δt < 250 s (4.17 min, dotted portion of the red line) as this is a too short timeframe for at least one IgG out of 39 to impinge at the gate in 100 μl. The ability of **Equation 5** to model also the responses as a function of the incubation time confirms the validity of a model based on the Brownian motion here discussed and confirms that with an incubation time of 9.2 minutes at least one-single molecule binding gives origin to a signal at the LOD level. This clearly proves that in 100 μl out of 8 ± 3 IgG molecules ($N_{LOD}$ in 100 μl) and in only in 9.2 min ($\Delta t_{LOD}$) at least one IgG reaches by pure diffusion the millimeter-wide gate. This demonstrates that the diffusion barrier issue does not apply when a millimeter wide interface is used to assay a volume of 100 μl or 1 ml, so that a SiMoT device can sense down to a LOD of 130 zM or lower depending on the measured level of the LOD. As an instance with a LOD level of 11.6%,[16] a LOD of 68 zM could have been reached. This also implies that a single IgG protein (footprint ~ $10^2$ nm$^2$ = $10^{-12}$ cm$^2$) can be successfully detected at the 0.2 cm$^2$ gate detecting surface, despite the $10^{11}$ orders of magnitude difference. Indeed, it is an ingrained belief that, the negligible footprint of a single molecule onto a many orders of magnitudes larger interface, returns a weak signal falling under the noise level. Experimental data prove this statement wrong. As an instance, some cells,[30],[31],[32], can detect a single photon, or a single chemoattractant, *via* their surface packed with capturing recognition elements. Indeed, amplification effects must be in place because a footprint of a single molecule (nanometric in size) is at least $10^8$ times smaller than the area of a cell surface (0.01-0.10 mm in size). Such occurrences, that are seen both in cells and FET-biosensors necessarily call for amplification effects discussed elsewhere.[19],[20],[16]

## 4. Conclusion

FET bioelectronic sensors comprising a millimeter-wide detecting gate covered by trillions of capturing elements such as antibodies, were proven capable to detect proteins at ultralow concentrations in the timeframe of minutes by several groups working on different technologies. Particularly relevant in this scenario is the SiMoT technology that has enabled single-molecule detection of proteins and genomic markers at the tens of zeptomolar concentration level, after an incubation time of only 10 minutes. These bioelectronic sensors can be fabricated by scalable large-area and cost-effective technologies and require no pretreatment or preparation of the sample to be assayed. Hence, they hold a tremendous potential in ultimately sensitive and fast detection of markers or even pathogens.

The sensing mechanism of this innovative approach is still under scrutiny and the present work adds a critically important piece of information: a single-molecule out of few, acting as a Browning particle diffusing according to the Einstein's diffusion-theory, can impinge on a large-area gate, functionalized with trillions recognition elements, within ten minutes. This is demonstrated by modelling the experimental data gathered with EG-FET SiMoT devices with a very simple expression (**Equation 5**) based on the Brownian theory. The acquired data, involves both the sensor response measured as a function of the volumes assayed at different concentrations (dose curves) and as a function of the incubation time. The data are very well reproduced by the same equation. Relevantly, the model also reveals that the fast spinning of the diffusing antigen enables also to quickly find the right orientation to bind to one capturing element independently of its orientation.

This work demonstrates that the *diffusion-barrier* issue, impairing the use of a single-molecule detection at a nanometric interface to assay solution with concentrations below picomolar, does not apply when the same experiment is conducted with a FET bioelectronic sensor comprising a micrometric or a millimetric wide detecting interface. It this case a single-molecule can be detected within few minutes in a 100 µl solution with a concentration down to few tens of zeptomolar.

## 5. Materials and Methods

*Materials*: The organic semiconductor channel material is a poly(3-hexylthiophene-2,5-diyl), P3HT (regioregularity > 99%), with an average molecular weight of 17.5 kDa (g mol$^{-1}$), used with no further purification. 3-mercaptopropionic acid (3-MPA) and 11-mercaptoundecanoic acid (11-MUA), 1-ethyl-3-(3-dimethylaminopropyl)-carbodiimide (EDC), N-hydroxysulfosuccinimide sodium salt (sulfo-NHS) and K4[Fe CN)6]·3H2O (98.5%) purchased from Sigma–Aldrich were also used with no further purification. The anti-Human Immunoglobulin G (anti-IgG, Sigma–Aldrich Product No. I2136) is a polyclonal antibody (molecular weight ~144 kDa) while the human IgG (~150 kDa) affinity ligand were extracted from human serum. Purified human IgG, purchased from Sigma-Aldrich (Product No. I 2511), is produced by precipitation and gel filtration techniques using normal human serum from one healthy donor as the starting material, to prevent the presence of dimer fraction in the sampled solution. The source material has been tested and found negative for antibody to HIV, antibody to HCV and for HbsAg. Bovine serum albumin (BSA) has a molecular weight of 66 kDa. All the proteins were purchased from Sigma–Aldrich and readily used. Water (HPLC-grade, Sigma-Aldrich), potassium chloride (Fluka, puriss p.a.) and ethanol grade, puriss. p.a. assay, ≥ 99.8 %, were also used with no further purification. All the electrical characterization and sensing experiments have been performed by means of a Keithley 4200-SCS semiconductor characterization system in air at room temperature in a dark box. All data were treated using OriginPro 2018, while Brownian diffusion-based modelling was implemented by means of Wolfram Mathematica software.

*Preparation of the IgG standard solutions*: The IgG PBS solutions were prepared by a serial dilution process with the dilution factor given by: $c_1 \times V_1 = c_2 \times V_2$, where $c_1$ and $c_2$ are the ligand concentrations in stock and in the diluted solution respectively, while $V_1$ and $V_2$ are the volumes of the stock and of the diluted solution, too. As customary, in a serial dilution process the former dilution is the stock solution for subsequent dilution in the series. The nominal number of the IgG proteins (# IgG) at each concentration was estimated as $\# IgG = \frac{c}{V N_A}$, where c is the ligand concentration, V is the volume of the standard PBS solution in which the gate is incubated, ranging from 25 µl to 1 ml, and $N_A$ is the Avogadro number. The uncertainty associated with the sampling in the serial dilution can be estimated, according to the Poisson's distribution, as the square root of the expected number of IgG proteins corresponding to one standard deviation.

*SiMoT electrolyte gated FET fabrication:* The EG-FETs, schematically shown in **Figure 1a**, were fabricated on a silicon substrate, covered by a 300 nm-thick $SiO_2$ layer. Source (S) and drain (D) interdigitated electrodes were photo-lithographically defined on the substrate and covered by a thiophene based organic semiconductor[33] Electron-beam evaporated Au films (50 nm-thick) were deposited on an adhesion layer of Ti (5 nm-thick). The channel length, 5 µm, and the channel width, 10.560 µm, define an effective channel area of $5.3 \cdot 10^{-2}$ mm². A solution of P3HT (2.6 mg·ml⁻¹ in chlorobenzene, filtered through a 0.2 µm sieve) was spin-coated at $2 \cdot 10^3$ *r.p.m.* for 20 s on these electrodes and annealed at 90°C for 15 s. A polydimethylsiloxane well was glued around the interdigitated channel area and filled with 300 µl of deionized water (HPLC-grade) serving as gating medium.[34] This is the SiMoT EG-FET *measuring well* comprising also two gate (G) electrodes. They hold a circular area of ~ 0.2 cm² and a $r_g = 0.25$ mm and were fabricated on PEN foil substrates by shadow-mask lithography and e-beam evaporation of Ti/Au (5 nm / 50 nm) films. The gate area being *ca.* 10 times that of the channel assures that the current relative change is mostly due to the shift of the gate work-function, or equivalently, the threshold voltage.[16] The gate serving as *sensing gate* undergoes a biofunctionalization process (*vide infra*) to be covered by the capturing anti-IgG antibodies. The other gate, addressed as *reference gate*, is made of bare-gold and measures the current level in the FET channel at any stage of a sensing measurement.

*Gate bio-functionalization protocol:* The sensing gate electrode was biofunctionalized according to a protocol described elsewhere.[20],[16] It comprises a 3-MPA and 11-MUA mixed chemical self-assembled monolayer (SAM) activated with EDC-sulfoNHS chemistry to whom anti-IgG capturing proteins are covalently attached. The unreacted activated carboxylic groups are deactivated in ethanolamine. The protocol enables to reach a coverage of capturing anti-IgG of $6 \cdot 10^{11}$. BSA is also physisorbed to minimize non-specific binding. The binding properties of the IgG analyte to the anti-IgG capturing layer is independently assessed by mean of a surface plasmon resonance characterization (**Figure S4** in **SI – Section 6**).

*Sensing measurements:* The *reference* gate is always in the measuring well and enables to control the level of the current flowing in the P3HT channel at every stage of the sensing assay. The biofunctionalized sensing gate shaffles from the *incubation* and the *measuring* well and it was proven that this does not provoke a shift the measured FET current of more than few %.[16] Before proceeding with the sensing measurements, the source-drain FET current is stabilized recording subsequent repeated measurements of the EG-FET I - V transfer curve (I *vs.* the gate bias at a fixed source-drain bias of - 0.3 V) in the *measuring* well using the *reference* gold gate, until a stable current is measured for three subsequent cycles. The *sensing* gate is then incubated in a separate well addressed as *incubation* well (or it is exposed to a droplet of the solution to be assayed) of a given volume of phosphate-buffered saline (PBS, ionic strength of 162 mM and pH of 7.4) solution (at RT and in the dark) for 10 min. Afterwards the sensing gate is washed thoroughly with HPLC water, transferred in the EG-FET *measuring* well and a new cycle of transfer characteristics is registered. Upon measurement of a stable $I_0$ base line, the same sensing gate is removed from the measuring well and transferred back into the incubation well filled with a PBS standard-solutions of the IgG molecules dispersed in different volumes. The incubations are caried out also for different given timeframes. Specifically, after incubation in each of the PBS standard-solutions the gate is washed thoroughly with HPLC grade water to remove physisorbed proteins, and the I-V transfer curve is measured in the measuring well. The stabilized currents measured after incubation in each standard solution are addressed as the "I" signal at a given concentration. The $(-\Delta I/I_0) = -(I - I_0)/I_0$ is the electronic response at a given volume/incubation time and the relevant curves are obtained by plotting the data at the gate-bias value that maximizes the trans-conductance $\delta I/\delta V$ (falling generally in the - 0.3 V to - 0.4 V range), at all the investigated volumes/incubation-times. All the data points are averaged over two or three replicates and the reproducibility error is computed as one standard deviation.

**Supporting Information**
Supporting Information is available from the Wiley Online Library or from the author.

**Acknowledgements**

David Walt is acknowledged for useful discussions. SiMBiT - Single molecule bio-electronic smart system array for clinical testing (Grant agreement ID: MIUR PON grants e-DESIGN (ARS01_01158); PMGB (ARS01_01195); IDF SHARID (ARS01_01270) 824946), Academy of Finland projects #316881, #316883 ''Spatiotemporal control of Cell Functions'', #332106 "ProSiT - Protein Detection at the Single-Molecule Limit with a Self-powered Organic Transistor for HIV early diagnosis", Åbo Akademi University CoE "Bioelectronic activation of cell functions" and CSGI are acknowledged for partial financial support.

Received: ((will be filled in by the editorial staff))
Revised: ((will be filled in by the editorial staff))
Published online: ((will be filled in by the editorial staff))

**Table of Content**

A single-molecule out of few in 100µl diffusing according to Einstein's theory, can impinge and be detected by a large-area (0.2cm$^2$) gate within ten minutes. This is demonstrated by modelling the data gathered with a bioelectronic FET whose gate is functionalized with a highly packed (10$^{12}$/cm$^2$) layer of capturing-antibodies. A general equation reproduces the data measured at different incubation volume and times.

*Eleonora Macchia, Liberato De Caro, Fabrizio Torricelli, Cinzia Di Franco, Giuseppe Felice Mangiatordi, Gaetano Scamarcio and Luisa Torsi*

**Why a diffusing single-molecule can be detected in few minutes by a large capturing bioelectronic interface**

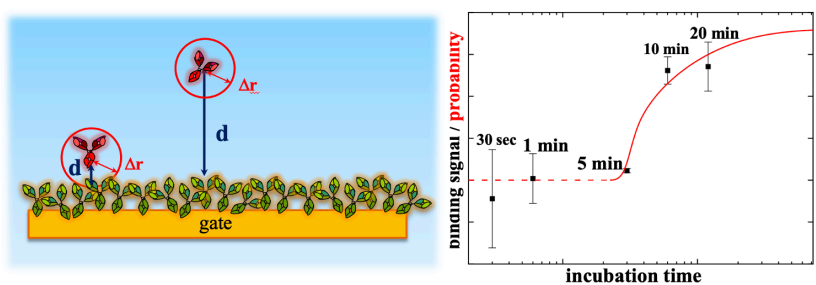

Supporting Information

**Why a diffusing single-molecule can be detected in few minutes by a large capturing bioelectronic interface**

*Eleonora Macchia, Liberato De Caro, Fabrizio Torricelli, Cinzia Di Franco, Giuseppe Felice Mangiatordi, Gaetano Scamarcio\* and Luisa Torsi\**

**Section 1: The 25 ml incubation volume case**

For the smallest volume assayed, a 25 µl, a droplet is deposited on the sensing gate. A schematic of the smallest volume assayed with the 25 µl droplet deposited directly on the gate surface, is also provided (**Figure S1a**). In this case, the radius of the droplet $r_d$ is measured to be 0.34 cm while that of the gate is $r_g = 0.25$ cm as shown in **Figure S1b**. **Equation 2** in the main text holds also in the case of the sensing gate exposed for 10 minutes to 25 µl droplet of the PBS standard solutions. **Figure S1a** shows that the whole gate area is covered by the droplet while the incubation system is schematically depicted in **Figure S1b.** The volume of a liquid droplet with contact angle $\theta$ and droplet radius $r_d$ is $V_d = \pi \cdot r_d^3 \cdot (2 - 3\cos\theta + \cos^3\theta) / (3 \cdot \sin^3\theta)$.[1] For $r_d \sim$ 3 mm a contact angle $\theta$ of about 60° is computed, accounting for the relatively high wettability of the densely packed anti-IgG surface. This is in line with a measured contact angle of 65° reported for a surface covered with a protein density of 0.25 µg/cm$^2$,[2] comparable to the $10^{12}$/cm$^2$ anti-IgGs covering the SiMoT gate surface. Also in this case, however, the portion of the solution that is close enough to the gate surface ($d \leq \Delta r$, see **Figure 1c**) to enable an antigen-antibody interaction within 600 s, can be approximated by a volume $V_{\Delta r}$ given by **Equation 2**, as $\theta \sim 90°$ and the cylindrical disk has a radius $r_d$ with a height of $\Delta r$.

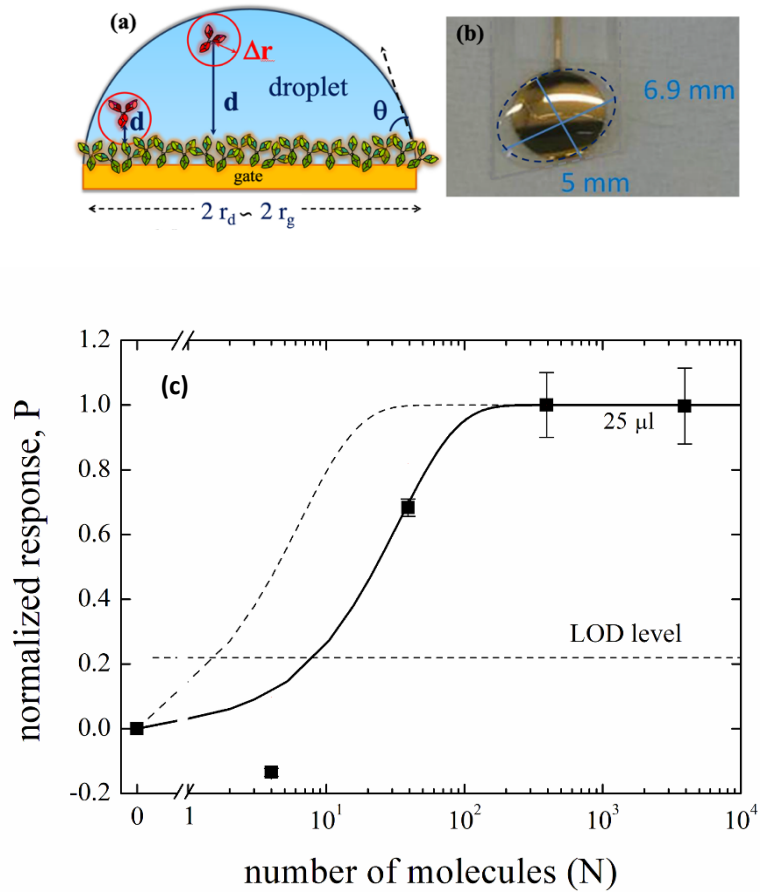

***Figure S1*** *– (a) Schematic cross-sectional view of the incubation step carried out in a droplet (volume = 25 µl) of radius $r_d$, forming a contact angle $\theta \sim 60°$ with the gate surface. (b) Top-view picture of the sensing gate covered by a 25 µl droplet during the incubation. (c) Normalized SiMoT EG-FET responses, $(\Delta I/I_0)/(\Delta I/I_0)^{sat}$ with $(-\Delta I/I_0)^{sat} = 0.65 \pm 0.12$, of anti-IgG functionalized gates incubated for 10 minutes (600 s) into a solution of IgG of 25 µl (black symbols) with the N number of IgG ligands ranging from $4 \pm 2$ to $3.92\ 10^4 \pm 2 \cdot 10^2$. Error bars are relevant to the reproducibility error indicated as one standard deviation over at least two replicates. On the y-axis the P probability function is also given, and the solid lines are the result of the modelling with the P function where the f function (Equation 3, main text) is multiplied by a* factor $\sim 0.19$. *The dotted line is the results of the modelling with the P probability function. The dotted horizontal line is the level of the LOD.*

The plot of the *P* probability function (**Equation 5**) results in the dotted curve given in **Figure S1c**. As it is apparent the P function is unable to reproduce the 25 µl data that are shifted towards higher N values. To fit the actual data the *f* function defined in the main text by the **Equation 3** (main text) needs to be multiplied by a factor of $\sim 0.19$. Assuming a constant $\Delta_R$ these results suggest a reduced effective surface in the case of the incubation by drop casting. Qualitatively, these findings may be explained by the well-known spontaneous dragging of liquid from the droplet's interior towards its edges (so called coffee-ring effect)[3],[4] that may preferentially distribute the diluted IgG molecules in the outer part of the droplet and reduce the effective interaction area of the gate.

**Section 2: Evaluation of the limit-of detection (LOD)**

The negative control experiment performed by exposing an anti-IgG gate to PBS standard solutions of Immunoglobulin M (IgM), defining the noise, is given in **Figure S2.**

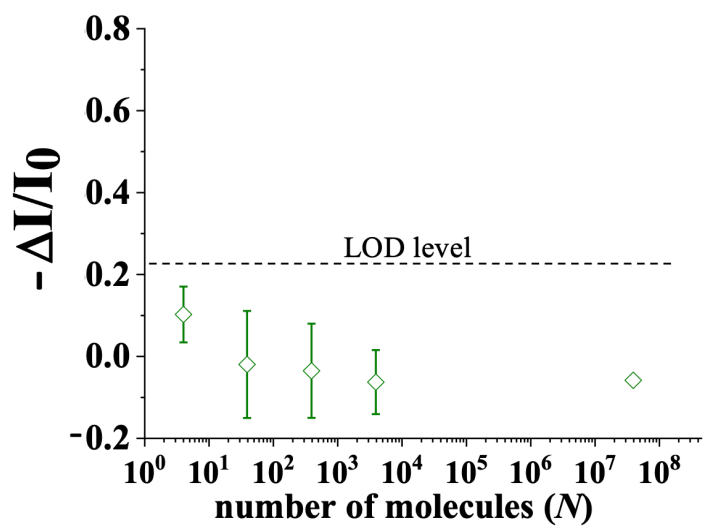

*Figure S2* - *Response of anti-IgG functionalized gates exposed to solutions of IgM molecules in PBS. Error bars are standard deviations over three replicates. The LOD level is computed as the average level of the signal (noise) plus three times its standard deviation.*

**Section 3: Further details on the Brownian motion theory**

The translational diffusion coefficient D is given by the Stokes-Einstein equation:

$$D = k_B \cdot T / (6 \pi \cdot \eta \cdot a) \qquad (S3.1)$$

with $k_B$ being the Boltzmann constant, T the absolute temperature, $\eta$ the solvent viscosity and *a* the hydrodynamic radius of the diffusing Brownian protein. Given the solute concentration c = 10 mM - KCl 2.7 mM and 137 mM NaCl - in the standard PBS solution (PH 7.4 and ionic-strength of 162 mM), we obtain a viscosity $\eta_s$:[5]

$$\eta_s/\eta = 1 + A\, c^{1/2} + B\, c \qquad (S3.2)$$

where η = 1·10$^{-2}$ g·cm$^{-1}$·s$^{-1}$ is the water viscosity at room temperature. In **Equation S3.2** A/(mol/l)$^{-1/2}$ and B/(mol/l)$^{-1}$ are equal to 0.0062 and 0.0793 for NaCl; these constants are equal to 0.0052 and -0.0140 for KCl. Inserting these values and the solute concentrations of the two salts in **Equation S3.2**, we obtain η$_s$ = 1.013 · 10$^{-2}$ g·cm$^{-1}$·s$^{-1}$, very close to the value for water. Taking $a$ = 5.51 ± 0.03 nm, the value for IgG monomers,[6] we obtain from Equation S1 a value D = 3.92 ± 0.02·10$^{-7}$ cm$^2$/s, in full agreement with photon-correlation spectroscopy data (3.89 ± 0.02·10$^{-7}$ cm$^2$/s).[6]

The dynamics of an IgG moving in a PBS (water) solution is described by the Langevin equation adapted to the case of an antigen - antibody interaction,[7] with the trajectory generated as a set of snapshots of the antigen position at each Δt time interval. The translational displacement **Δr** is given by: **Δr** = (k$_B$ T)$^{-1}$ · D · **F** · Δt + **R** where **F** is the force acting on the antigen before the step is taken, **R** is the random vector satisfying both < **R** > = 0 and < **R**$^2$ > = 6 D · Δt. The factor (k$_B$ T)$^{-1}$ · D models the damping effect generated by the friction with the molecules of the solvent. As in the case under study, the anti-IgG capturing antibodies are covalently attached at the gate surface while the IgG antigen diffuses in the solvent, the latter can be treated as a rigid body whose translational motion is referred to the position of the gate. For the diffusion to proceed it is necessary that, within Δt, the forces and the torques acting on the IgG antigen, as well as the gradient of any diffusion tensor, remain effectively constant. Hence, the motion should be considered only over timeframes exceeding the momentum relaxation time Δt$_r$ = ($m$ · D) / k$_B$ T with $m$ being the antigen mass. Plugging $m$ ~ (ρ · 4π · $a^3$) / 3 and **Equation S3.1** into Δt$_r$, results in:

$$\Delta t_r = 2a^2 \cdot \rho / (9 \cdot \eta) \tag{S3.3}$$

that falls in the picosecond range. This proves that the condition for the Brownian motion to hold, is met. It is straightforward to show that hydrodynamic interactions and gravitational effects can be neglected.

The Brownian motion give rise also to a rotational diffusion with a random rotational angle **W** satisfying <**W**> = 0 so that

$$<\mathbf{W}^2> = 6 D_R \Delta t. \tag{S3.4}$$

with a $D_R$, rotational diffusion coefficient being, according to the Einstein–Smoluchowski $D_R = k_b T/8\pi\eta a^3$. For an antibodies $D_R \sim 1$ MHz. If we put $(<\mathbf{W}^2>)^{1/2} = 4\pi$, $a = 5.51$ nm, from **Equation 2.4** we obtain that the time an IgG takes to span the whole solid angle $4\pi$, is $\Delta t_{4\pi} = 16\pi^2/6 D_R \sim 25$ μs.

**Section 4: Complete expression of the P probability function**

The complete expression of the probability function $P$ is provided in the following **Equation S4.1**:

$$P = \left[\frac{\pi}{2}\sqrt{6D\,\Delta t}\cdot r_g^2/V\right]\left\{1 + \left[1 - \frac{\pi}{2}\sqrt{6D\,\Delta t}\cdot r_g^2/V\right] + \left[1 - \frac{\pi}{2}\sqrt{6D\,\Delta t}\cdot r_g^2/V\right]^2 + \cdots \left[1 - \frac{\pi}{2}\sqrt{6D\,\Delta t}\cdot r_g^2/V\right]^{N-1}\right\} \tag{S4.1}$$

Details on the probability function P as function of the incubation volume encompassing a different number of molecules N is given in **Figure S3**.

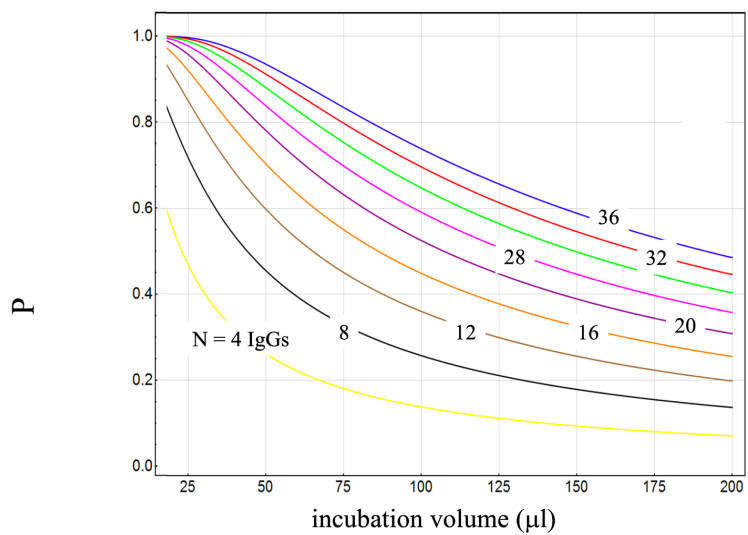

***Figure S3** - Probability P that in an incubation time of 600 s at least one of N IgGs, with N ranging between 4 and 36, will be close enough to the gate so as, during its random Brownian motion, the binding with an anti-IgG will occur. The plot starts from 16 μl, corresponding to the minimum volume needed to cover the gate.*

**Table S1:** values for all the parameters present in **P**

| Physical parameter | description | value/origin |
|---|---|---|
| $r_g$ | Gate radius<br>*Measured* | 0.25 cm |
| $\Delta t$ | incubation time<br>*Measured* | 30 s, 1 min, 5 min, 10 min and 20 min |
| N | number of particles in the assayed volume V<br>$N = [C] (V \cdot N_A)$ with $N_A$ = Avogadro's number and [C] = molar concentration<br>*Measured* | from $4 \pm 2$ to $3.92 \cdot 10^7 \pm 6 \cdot 10^3$ |
| V | assayed incubation volume<br>*Measured* | 25 µl, 100 µl, 1 ml |
| T | Temperature<br>*Measured* | 24 °C |
| $K_B$ | Boltzmann constant | $1.38064852 \cdot 10^{-23}$ m² kg s⁻² K⁻¹ |
| D | Diffusion coefficient of IgG monomer (T. Jøssang, J. Feder, E. Rosenqvist, J. Protein Chem. **1988**, 7, 165)<br>*Measured* | $3.89 \cdot 10^{-7}$ cm²/s |

**Section 5: Chi-squared test**

The chi-squared ($\chi^2$) test obtained by fitting experimental data in **Figure 2** with **the probability function P**, by considering just the first experimental point for the saturation plateau (k = degree of freedom; p = probability > $\chi^2$)[8], are:

$V = 25\ \mu l,\ \chi^2 = 0.073\ (k = 2),\ p = 0.96;$

$V = 100\ \mu l,\ \chi^2 = 0.051\ (k = 2),\ p = 0.97;$

$V = 1\ ml,\ \chi^2 = 0.093\ (k = 2),\ p = 0.955;$

The chi-squared test for the data in **Figure 3** modelled also with **function P** gives:

$\chi^2 = 0.0345\ (k = 1),\ p = 0.85.$

Since *p > 0.95* in all the four sets of data at different incubation volumes, while p is larger than 0.85 when fitting the response at different incubation times, there is a high chance (95% for the data of **Figure 2** and 85% for the data of **Figure 3**) that the null hypothesis is correct, *i.e.,* that there is no difference between the observed and theoretical (expected) values predicted by the modeling.

**Section 6: Surface Plasmon Resonance study of IgG binding to anti-IgG**

The Surface Plasmon Resonance (SPR) technique has been used to characterize the gold surface modified with the bio-recognition element, anti-IgG, in terms of number of antibodies immobilized and the capturing efficacy against the IgG affinity ligand. A Multi-Parameter SPR (MP-SPR) Navi 200-L apparatus in the Kretschmann configuration was used. An Au coated (~50 nm) SPR slides (BioNavis Ltd) comprising a chromium adhesion layer (~2 nm) served as semi-transparent SPR substrate. The SPR slide holding an area of 0.42 cm$^2$ was inspected in two different spots by two laser sources, both set at 670 nm, to estimate the layer homogeneity. The SPR slide was cleaned in a $NH_4OH/ H_2O_2$ aqueous solution (1:1:5 v/v) at 80-90°C for 10 min, and treated in an ozone cleaner for 10 min. The gold surface was modified *ex-situ* with a mixed self-assembled monolayer (SAM) of alkylthiols: 11-mercaptoundecanoic acid (11MUA) and 3-mercaptopropionic acid (3MPA) in molar ratio 1:10 in ethanol holding a concentration of 10 mM. The sample, immersed in the thiol solution, was left overnight in nitrogen atmosphere at room temperature. Afterwards, the slide was rinsed in ethanol and mounted in the SPR sample holder. The modified SPR slide was further bio-functionalized in the SPR apparatus, thus the real-time anchoring of antibodies on the SAM was monitored *in-situ* as depicted in **Figure S4a**. To achieve the bio-conjugation of antibodies on the SAM, the established coupling method with 1-Ethyl-3-(3-dimethylamino-propyl)carbodiimide (EDC) and N-hydroxysulfosuccinimide sodium salt (NHSS) was used. To this aim, the SAM surface was kept in contact with the aqueous solution of EDC/NHSS (0.2/0.05 M) for 20 minutes. The carboxylic terminal groups of the chemical SAM are converted into intermediate reactive species (NHSS, N-hydroxysulfosuccinimide esters), reacting with the amine groups of the antibody, anti-IgG, achieving its covalent coupling. This was carried out by static injection of 100 μL of the phosphate buffer solution (PBS, ionic strength $i_s$ 163 mM and pH 7.4) of the anti-IgG capturing antibodies (100 μg/mL) in the SPR cell. Then, the ethanolamine saturated solution (EA, at concentration 1 M) is injected to deactivate the unreacted esters in an inactive hydroxyethyl amide. Finally, to cover possible voids on the SAM and to prevent non-specific binding, a BSA solution 100 µg/mL in PBS was used.

The surface coverage of anti-IgG covalently bound on the SAM was quantitatively assed by means the de Feijter's equation (**Equation S6.1**):

$$\Gamma = d \cdot (n - n_0) \cdot (dn/dC)^{-1} \qquad (S6.1)$$

where $\Gamma$, expressed in ng cm$^{-2}$, is the surface coverage, d the thickness of the biolayer deposited on the gold surface, n-n$_0$ is the difference between the refractive index of the adlayer and the one of the bulk medium, and dn/dC is the specific refractivity of the adsorbed biolayer. Deriving this further to consider the instrument response, it returns **Equation S6.2**:

$$(n - n_0) = \Delta\theta_{SPR} \cdot k \qquad (S6.2)$$

where k is the wavelength dependent sensitivity coefficient, and $\Delta\theta_{SPR}$ is the experimental angular shift. For laser beams with $\lambda$ = 670 nm and a thin layers (d < 100 nm), the following approximations hold true: (i) dn/dC ≈ 0.182 cm$^3$ g$^{-1}$, (ii) k·d ≈ 1.0·10$^{-7}$ cm deg.[9] Therefore, under these assumptions and by substitution of **Equation S6.1** in **Equation S6.2**, **Equation S6.3** is derived to estimate the surface coverage $\Gamma$ using the experimental angular shift, being

$$\Gamma = \Delta\theta_{SPR} \cdot 550 \; [ng/cm^2] \qquad (S6.3).$$

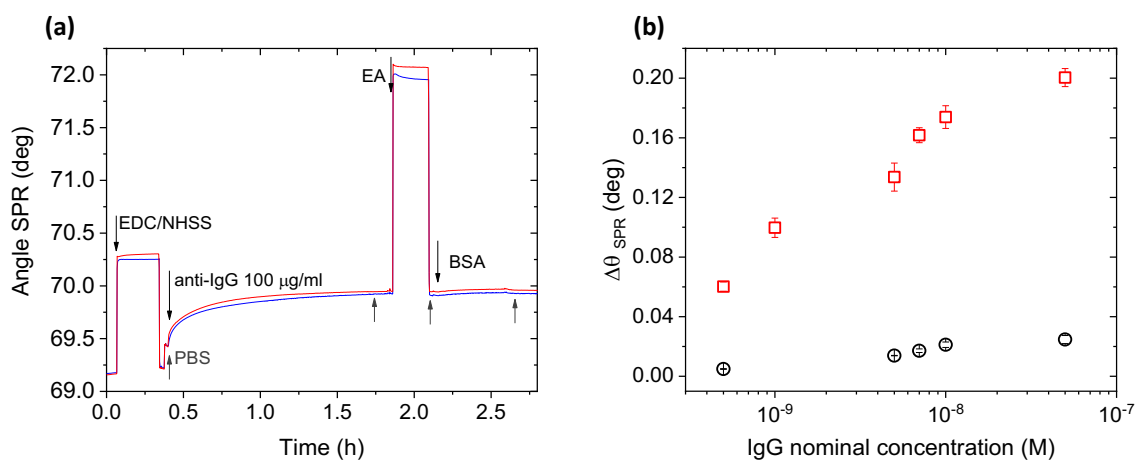

***Figure S4*** *– (a) SPR sensogram (plasmon peak angle vs. time) of the anti-IgG covalent immobilization through mixed-SAM on the gold SPR slide. Red and blue curves are relevant to two points of the slide surface simultaneously inspected. (b) SPR dose-curves ($\Delta\theta_{SPR}$ vs.*

*[IgG] nominal concentrations on semi-log scale) for the IgG binding on anti-IgG (red squares) and on BSA (black circles) covalently bound on the chemical SAM.*

Thus, in **Figure S4b** (red squares) the dose-curve for the assayed IgG is reported as $\Delta\theta_{SPR}$ *vs.* [IgG] nominal concentrations (semi-log scale). A control experiment was performed by using the BSA instead of anti-IgG as capturing bio-recognition element. Therefore, the SAM was modified with a BSA solution in PBS at concentration 100 µg/mL, following the same protocol previously described for the anti-IgG anchoring. In **Figure S4b** (black circles) the response of the control sample upon the exposure of IgG is reported as well as $\Delta\theta_{SPR}$ *vs.* [IgG] nominal concentrations. An anti-IgG surface coverage of 288 ± 17 ng cm$^{-2}$, corresponding to (1.16 ± 0.3) ·10$^{12}$ molecules·cm$^{-2}$, has been registered 1 hour and 30 minutes after the PBS washing step. The binding efficacy of the bio-recognition elements immobilized on the SAM was tested against IgG. The analysis was carried out by recording the baseline in PBS and injecting IgG solutions in PBS at different concentrations, ranging from 5·10$^{-10}$ M to 5·10$^{-8}$ M. The angular shift, $\Delta\theta_{SPR}$, was calculated for each concentration as the difference between the equilibrium value, after rinsing with PBS, and the initial baseline.